\documentclass[11pt,a4paper]{article}
\usepackage{jheppub}
\usepackage{axodraw}
\usepackage{epsfig}
\usepackage{xspace}
\usepackage{graphics}
\usepackage[utf8]{inputenc}
\usepackage{relsize}

\newcommand{\atlas}{ATLAS\xspace}
\newcommand{\ariadne}{A{\smaller RIADNE}\xspace}
\newcommand{\alpgen}{A{\smaller LPGEN}\xspace}
\newcommand{\pythia}{P{\smaller YTHIA}\xspace}
\newcommand{\sherpa}{S{\smaller HERPA}\xspace}
\newcommand{\alphas}{\ensuremath{\alpha_s}}
\newcommand{\HEJ}{\emph{HEJ}\xspace}

\newcommand{\done}[1]{}
\newcommand{\refeq}[1]{Eq.~\eqref{#1}}
\newcommand{\Ca}{\ensuremath{C_{\!A}}\xspace}

\newcommand{\Nc}{\ensuremath{N_{\!C}}\xspace}
\newcommand{\as}{\ensuremath{\alpha_s}\xspace}
\newcommand{\gs}{\ensuremath{g}}
\newcommand{\eg}{\emph{e.g.}\xspace}

\subheader{
  \begin{flushright}\sffamily
    CERN-PH-TH-2011-072\\
    CP3-Origins-2011-14\\
    Edinburgh 2011/16\\
    LU-TP 11-15\\
    MCnet-11-12
  \end{flushright}
}

\title{A Parton Shower for High Energy Jets\footnotetext[1]{Work supported in
    parts by the EU Marie Curie RTN MCnet (MRTN-CT-2006-035606),
    the Swedish research council (contracts 621-2008-4252 and 621-2009-4076) and
    the UK Science and Technology Facilities Council (STFC).}}

\author[a]{Jeppe R.~Andersen,}
\author[b,c]{Leif Lönnblad}
\author[d]{and Jennifer M.~Smillie}
\emailAdd{andersen@cp3.sdu.dk}
\emailAdd{Leif.Lonnblad@thep.lu.se}
\emailAdd{j.m.smillie@ed.ac.uk}
\affiliation[a]{{CP}$^{ \bf 3}${-Origins}, University of Southern Denmark, Campusvej 55, DK-5230 Odense M, Denmark.}
\affiliation[b]{CERN Theory Department, CH-1211 Geneva, Switzerland}
\affiliation[c]{Dept.~of Astronomy and Theoretical Physics, Lund University,
  S\"olvegatan 14A, SE-223 63 Lund, Sweden}
\affiliation[d]{School of Physics and Astronomy, University of Edinburgh,
  Mayfield Road, Edinburgh EH9 3JZ, UK}
  \abstract{We present a method to match the multi-parton states
    generated by the High Energy Jets Monte Carlo with parton showers
    generated by the \ariadne program using the colour dipole
    model. The High Energy Jets program already includes a full
    resummation of soft divergences. Hence, in the matching it
    is important that the corresponding divergences in the parton
    shower are subtracted, keeping only the collinear parts. We present
    a novel, shower-independent method for achieving this, enabling us to
    generate fully exclusive and hadronized events with multiple hard jets,
    in hadronic collisions. We discuss in detail the arising description of
    the soft, collinear and hard regions by examples in pure QCD
    jet-production.}
\begin{document}
\maketitle 
\sloppy
 
\section{Introduction}
\label{sec:intro}
The CERN Large Hadron Collider is testing not only several suggested
extensions of the Standard Model, but is simultaneously testing the
theoretical description of particle collisions within the Standard
Model. With the increased energy at the LHC over earlier colliders,
the accurate theoretical description of several observables will
necessitate the inclusion of several additional jets of hardness similar to
that of the relevant lowest-order process being studied.

The all-order perturbative description included in the general-purpose
Monte Carlo
programs~\cite{Sjostrand:2007gs,Marchesini:1988cf,Bahr:2008pv,Gleisberg:2008ta}
relies on properties of soft- and collinear radiation from the
lowest-order hard process, and therefore underestimates the amount of
radiation of similar hardness. The shower-description of radiative
corrections can be corrected above a chosen merging-scale with the
full tree-level matrix elements with, \eg, the
CKKW(-L)~\cite{Catani:2001cc,Lonnblad:1992tz} or MLM~\cite{Mangano:2001xp}
approach. The multiplicity, to which the tree-level matrix
elements are required to be evaluated, will depend on the chosen
merging scale. A low merging scale will correct the parton shower in
much of phase space, but the multiplicity to which the tree-level matrix
elements must be evaluated is then very large (and in practice the maximum
multiplicity is limited).

Several alternative approaches have been developed in order to respond
to the need for the description of semi-hard emissions to all orders,
not just those which can be reached by matching from the
shower-formalism. Some of these approaches, like
Cascade~\cite{Jung:2000hk,Jung:2010si} and
\ariadne~\cite{Lonnblad:1992tz}, originate from the study of specific
phase space regions at earlier colliders, where the semi-hard
radiative corrections were already deemed relevant. In the formulation
of Cascade, the semi-hard emissions are generated from the
PDF-evolution according to the CCFM
equation~\cite{Ciafaloni:1987ur,Catani:1989yc,Catani:1989sg,Marchesini:1994wr},
while \ariadne implements the \emph{Colour Dipole
  Model}~\cite{Gustafson:1986db,Gustafson:1987rq,Andersson:1988gp,Lonnblad:1994wk}
of the evolutions of colour dipoles after the hard scattering.

The current study uses the new approach of \emph{High Energy Jets}
(\HEJ)~\cite{Andersen:2009nu,Andersen:2009he,Andersen:2011hs}, which is
built on standard collinear factorisation between hard scattering
matrix elements and PDFs, but calculates the hard scattering matrix
elements to all orders. This is achieved within an approximation,
which becomes exact in the limit of large invariant mass between all
particles. This is the exact opposite limit of collinear
emissions. The close similarity between the formulation of the
resummation in \HEJ and the normal perturbative expansion of a fixed-order
calculation allows for a simple procedure\cite{Andersen:2011hs} for matching
the \HEJ resummation to full tree-level accuracy, similar to the merging of a
parton shower and matrix elements in a CKKW-L or MLM merging procedure.

The multi-jet predictions from \HEJ are in the form of partonic final states,
with no collinear enhancement of radiation. While \HEJ describes the jet count
and topology, jet shapes are completely ignored. The jet cones are mostly
empty, except for the single parton taking up all the jet momentum, not
unlike the situation of a tree-level generator.

In order to arrive at a more realistic description of the final state of the
particle scattering, first a resummation of the (soft\footnote{We will in fact
  see later that much of the soft radiation is already resummed in \HEJ.}  and)
collinear emissions of a parton shower is necessary, in order to secondly add a
hadronisation step. While the description of jet profiles within \HEJ is not
unlike the situation at tree-level, the challenge of matching the description to
a shower is completely unlike that at tree-level solved in the approach of
CKKW~\cite{Catani:2001cc,Lonnblad:1992tz}, MLM~\cite{Mangano:2001xp} or \textsc{Vincia}\cite{Giele:2007di}. This is
because \HEJ is summing its own tower of real corrections to all orders
and, on top of that, also includes the leading virtual corrections in the
limit of hard, wide angle emissions. The
challenge of matching \HEJ with a shower is to avoid double counting between the
two all-order approaches.

We will present a shower-independent subtraction algorithm for matching \HEJ
and a parton shower. Furthermore, this paper contains a study of the effects
of the parton shower and hadronisation on the predictions arising from
\HEJ. This is obtained for an implementation of the specific evolution of the
final state according to \ariadne, with the string hadronisation as
implemented in \pythia\cite{Sjostrand:2000wi,Sjostrand:2003wg}. In
Section~\ref{sec:hej} we present the ingredients of \HEJ necessary for the
further discussion, and in Section~\ref{sec:ariadne} we do the same for
\ariadne. Section~\ref{sec:matching} presents the matching of the two
all-order perturbative approaches. Section~\ref{sec:results} discusses the
impact of the addition of the shower and hadronisation on various classes of
observables, both some very sensitive to the description of collinear
radiation (the shower profile), and some which will turn out to be modified
only slightly (\eg~vetos of hard jets). The addition of a parton shower on
top of \HEJ should extend the validity of the prediction to regions with a
large ratio also of transverse scales.

Finally, in Section \ref{sec:outlook} we present our conclusions and
outline some future improvements of our matching procedure.

\section{The High Energy Jets Monte Carlo}
\label{sec:hej}
The \emph{High Energy Jets (\HEJ)}
framework~\cite{Andersen:2009nu,Andersen:2009he,Andersen:2011hs}
provides a perturbative approximation to the hard scattering matrix
elements to jet production to any order in the coupling, which is
exact in the limit of large invariant mass between all particles. The
formalism is inspired by the high energy factorisation of matrix
elements (as pioneered by
BFKL~\cite{Fadin:1975cb,Kuraev:1976ge,Kuraev:1977fs,Balitsky:1978ic}),
in that only certain partonic configurations are described (namely
those leading in the limit of large invariant mass between all
partons, also denoted Multi-Regge Kinematics --- \emph{MRK}). Within
the framework of BFKL, a number of kinematic approximations are
applied in order to cast the cross section in the form of a
two-dimensional integral equation. This also entails a number of
approximations to the phase space. These approximations are avoided
within \HEJ. The advances in computing power allow the calculation of
the cross section as an explicit integration over matrix elements of
each multiplicity, with the inclusion of (an approximation to the)
virtual corrections, and a simple organisation of the cancellation of
IR singularities between real and virtual corrections. This allows for
the construction of an approximation, which retains the logarithmic
accuracy of the BFKL approach, but is more accurate when compared to
the full QCD amplitudes in the phase space regions relevant for the
LHC. The details are discussed in
Ref.~\cite{Andersen:2009nu,Andersen:2009he,Andersen:2011hs}; here we
will briefly repeat the discussion of the points which are necessary
for constructing the matching to a parton shower.

The all-order treatment in \HEJ starts with the approximation to the
tree-level scattering amplitude for the scattering process with
flavours $f_1 f_2\to f_1 g\cdots gf_2$, where the final state
particles are listed according to their rapidity, and $f_1, f_2$ can
be quarks, anti-quarks or gluons. We will call these states
\emph{FKL}-configurations
(Fadin-Kuraev-Lipatov~\cite{Kuraev:1976ge}). The scattering amplitude
is approximated at lowest order by the following
expression~\cite{Andersen:2011hs}
\begin{align}
  \label{eq:multijetVs}
  \begin{split}
    \left|\overline{\mathcal{M}}^t_{f_1 f_2\to f_1g\ldots gf_2}\right|^2\ =\ &\frac 1 {4\
      (\Nc^2-1)}\ \left\|S_{qQ\to qQ}\right\|^2\\
    &\cdot\ \left(g^2\ K_{f_1}\ \frac 1 {t_1}\right) \cdot\ \left(g^2\ K_{f_2}\ \frac 1
      {t_{n-1}}\right)\\
    & \cdot \prod_{i=1}^{n-2} \left( \frac{-g^2 C_A}{t_it_{i+1}}\
      V^\mu(q_i,q_{i+1})V_\mu(q_i,q_{i+1}) \right),
  \end{split}
\end{align}
where $\left\|S_{f_1 f_2\to f_1 f_2}\right\|^2$ indicates the square of pure
current-current scattering, and $K_{f_1}, K_{f_2}$ are flavour-dependent
colour-factors (which can depend also on the momentum of the particles of
each flavour $f_1, f_2$, see Ref.\cite{Andersen:2011hs} for more
details). The use of flavour-dependent colour factors allows one to display
explicitly the complete factorisation at tree-level of all QCD processes $f_1
f_2 \to f_1 f_2$ into contractions of normal currents over a $t$-channel
pole. $g^2=4\pi \as$ is the QCD coupling, and $t_i$ is the square of the
local $t$-channel momentum, $t=q_i^2$, $q_i=p_a-\sum_{j=1}^i p_j$, where
$p_a$ momentum of the incoming parton of negative $z$-momentum, and the
momenta $p_i$ of the outgoing partons are ordered with increasing rapidity.

The effective vertex for emissions of gluons takes the
form~\cite{Andersen:2009nu}
\begin{align}
  \label{eq:GenEmissionV}
  \begin{split}
  V^\rho(q_i,q_{i+1})=&-(q_i+q_{i+1})^\rho \\
  &+ \frac{p_A^\rho}{2} \left( \frac{q_i^2}{p_{i+1}\cdot p_A} +
  \frac{p_{i+1}\cdot p_B}{p_A\cdot p_B} + \frac{p_{i+1}\cdot p_n}{p_A\cdot p_n}\right) +
p_A \leftrightarrow p_1 \\ 
  &- \frac{p_B^\rho}{2} \left( \frac{q_{i+1}^2}{p_{i+1} \cdot p_B} + \frac{p_{i+1}\cdot
      p_A}{p_B\cdot p_A} + \frac{p_{i+1}\cdot p_1}{p_B\cdot p_1} \right) - p_B
  \leftrightarrow p_n.
  \end{split}
\end{align}
This form of the effective vertex is fully gauge invariant; the Ward
Identity, $p_j\cdot V=0$ ($j=2,...,n-1$) can easily be checked, and is
valid for all momenta $p_j$ (i.e.~not just in the \emph{MRK}-limit). This allows for a meaningful approximation to the
scattering amplitude to be constructed.

The virtual corrections are approximated with the \emph{Lipatov
  ansatz}\cite{Balitsky:1978ic} for the $t$-channel gluon propagators (see
Ref.\cite{Andersen:2009nu} for more details). This is obtained by the simple
replacement in Eq.~(\ref{eq:multijetVs}) of
\begin{align}
  \label{eq:LipatovAnsatz} \frac 1 {t_i}\ \to\ \frac 1 {t_i}\ \exp\left[\hat
\alpha (q_i)(y_{i-1}-y_i)\right]
\end{align} with
\begin{align} 
  \hat{\alpha}(q_i)&=-\gs^2\ \Ca\
  \frac{\Gamma(1-\varepsilon)}{(4\pi)^{2+\varepsilon}}\frac 2
  \varepsilon\left({\bf q}_i^2/\mu^2\right)^\varepsilon\label{eq:ahatdimreg},
\end{align} 
where $\hat \alpha$ is regulated in $D=4+2\varepsilon$ dimensions and
$\mathbf{q}_i^2$ is the Euclidean square of the transverse components of
$q_i$. The cancellation of the poles in $\varepsilon$ between the real and
virtual corrections is organised with a mix of a subtraction and a phase space
slicing (basically limiting the phase space region of the subtraction terms),
such that the regulated matrix elements used in the resummation of \emph{HEJ}
are given by\cite{Andersen:2011hs}
\begin{align}
  \label{eq:MHEJ}
  \begin{split}
    \overline{\left|\mathcal{M}_{\rm HEJ}^{\mathrm{reg}, f_1 f_2\to f_1 g
          \cdots g f_2}(\{ p_i\})\right|}^2 = \ &\frac 1 {4\
       (\Nc^2-1)}\ \left\|S_{f_1 f_2\to f_1 f_2}\right\|^2\\
     &\cdot\ \left(g^2\ K_{f_1}\ \frac 1 {t_1}\right) \cdot\ \left(g^2\ K_{f_2}\ \frac 1
       {t_{n-1}}\right)\\
     & \cdot \prod_{i=1}^{n-2} \left( {g^2 C_A}\
       \left(\frac {-1}{t_it_{i+1}} V^\mu(q_i,q_{i+1})V_\mu(q_i,q_{i+1}) -
         \frac{4}{\mathbf{p}_i^2}\ \theta\left(\mathbf{p}_i^2<\lambda^ 2\right)\right)\right)\\
     & \cdot \prod_{j=1}^{n-1} \exp\left[\omega^0(q_j,\lambda)(y_{j-1}-y_j)\right], \\
     \omega^0(q_j,\lambda)=\ &-\frac{\alpha_s \Ca}{\pi} \log\frac{{\bf q}_j^2}{\lambda^2}.
  \end{split}
\end{align}

The all-order dijet cross section is then simply calculated as the phase
space integral over any number of gluon emissions from the initial scattering
$f_1 f_2 \to f_1 f_2$. Matching to high-multiplicity tree-level matrix
elements is obtained by reweighting the event with the ratio of the square of
the tree-level matrix element (evaluated using MadGraph\cite{Alwall:2007st})
and the approximation to this in Eq.~(\ref{eq:multijetVs}), both evaluated on
a set of momenta derived from the hard jets only (in order to reduce the
multiplicity, and therefore not to exhaust the limited number of available
full tree-level matrix elements too quickly). This
procedure is summarised in the following formula
\begin{align}
  \begin{split}
    \label{eq:resumdijetFKLmatched}
    \sigma_{2j}^\mathrm{resum, match}=&\sum_{f_1, f_2}\ \sum_{n=2}^\infty\
    \prod_{i=1}^n\left(\int_{p_{i\perp}=0}^{p_{i\perp}=\infty}
      \frac{\mathrm{d}^2\mathbf{p}_{i\perp}}{(2\pi)^3}\ 
      \int \frac{\mathrm{d} y_i}{2}
    \right)\
    \frac{\overline{|\mathcal{M}_{\mathrm{HEJ}}^{f_1 f_2\to f_1 g\cdots gf_2}(\{ p_i\})|}^2}{\hat s^2} \\
    &\times\ \sum_m \mathcal{O}_{mj}^e(\{p_i\})\ w_{m-\mathrm{jet}}\\
    &\times\ \ x_a f_{A,f_1}(x_a, Q_a)\ x_2 f_{B,f_2}(x_b, Q_b)\ (2\pi)^4\ \delta^2\!\!\left(\sum_{i=1}^n
      \mathbf{p}_{i\perp}\right )\ \mathcal{O}_{2j}(\{p_i\}),
  \end{split}
\end{align}
where $n$ is the partonic multiplicity of the final state. The second line in
Eq.~(\ref{eq:resumdijetFKLmatched}) describes the matching of \HEJ to
high-multiplicity tree-level by a reweighting of the exclusive $m$-jet rate
projected with the operator $\mathcal{O}^e_{mj}$ with the ratios of the
relevant tree-level matrix elements. The last line describes the inclusion of
PDFs and the momentum-conserving delta-functional. We define the inclusive
two-jet operator $\mathcal{O}_{2j}$ to return one if the final state contains
at least two hard ($p_\perp>35$~GeV) jets; in the current study, we use the
anti-kt algorithm with an $R=0.6$, and the $E$-recombination
scheme. Furthermore, we require that the extremal partons from HEJ are
members of the extremal jets, in order to ensure that the partonic
configuration matches the situation for which the \HEJ resummation scheme was
developed.

The partonic configurations not conforming to the ordering described above
are included in \HEJ by simply adding the contributions order-by-order (again
using \texttt{MadGraph}~\cite{Alwall:2007st}), but no all-order summation is
performed of these non-\emph{FKL} configurations. In the current study, we
will focus on the \emph{FKL}-configurations, since this is where special
all-order attention is needed in order to avoid double counting in the subsequent
shower. The non-FKL configurations could possibly be added to the combined
\HEJ\!\!\!+shower sample through \eg a vetoed CKKW-L-procedure (vetoing any
\emph{FKL}-configuration which might arise).

The matching of \HEJ to high-multiplicity tree-level accuracy is currently
performed with up to four jets in the final state, limited by the time taken
to evaluate the full expressions.  Importantly, the description in \HEJ goes
beyond approximating leading-order high-multiplicity matrix elements.  As discussed, the \emph{Lipatov
  ansatz}~\cite{Balitsky:1978ic} is used to give an approximation to the
virtual corrections at all orders in addition. This resums to all-orders the
leading logarithmic virtual corrections to the $t$-channel poles.

By construction, the \HEJ framework therefore provides a description directed
particularly at hard, wide-angle QCD radiation. The transverse momenta of
gluons emitted in-between (in rapidity) the two extremal partons can take on
any value, however the subtraction suppresses the contribution from emissions
with a transverse momentum less than $\lambda$. We note that although
Eq.~(\ref{eq:MHEJ}) contains the exponential of the logarithm of a momentum,
it is not directly related to a Sudakov factor of a normal parton shower ---
firstly, the logarithm is not of the emitted momentum, secondly, the
resummation is not
formulated as a unitary evolution (i.e.~of constant total cross section). In
order to perform a matching to a parton shower, we have to \emph{define} the
relation between the emissions of \HEJ and those of the shower. Since the
emissions of \HEJ populate all of phase space (in-between in rapidity of the
scatterers of flavour $f_1$ and $f_2$), we do not want to define specific
regions to populate with \HEJ and with the shower. Rather, we will let both
formalisms populate their respective phase spaces, but define a subtraction
term for the shower Sudakov, such that double counting is avoided by reducing
the probability of a certain emission from the shower by the probability that
\HEJ had already performed the given emission.

A parton shower framework, such as \ariadne \cite{Lonnblad:1992tz}, is
necessary in order to evolve the partonic state of \HEJ to the state of
hadronisation, primarily by populating the partonic state with further soft
and collinear radiation.  As the shower, and also the subsequent string
hadronization, relies on having well-defined colour connection between
partons, we first need to briefly discuss how these are obtained from \HEJ.

\subsection{The Colour Connections of High Energy Jets}
\label{sec:colo-conn-high}
\begin{figure}
  \hspace{2cm}
  \begin{minipage}[b]{0.5cm}
    \begin{flushright}
      $a$

      \vspace{3.3cm}
      $b$

      \vspace{0.3cm}
    \end{flushright}
  \end{minipage}
  \epsfig{height=4.5cm,file=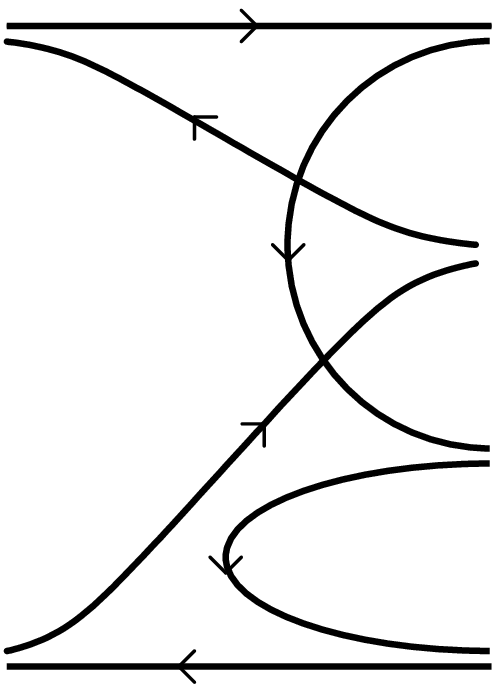}
  \hspace{2.5cm}
  \epsfig{height=4.5cm,file=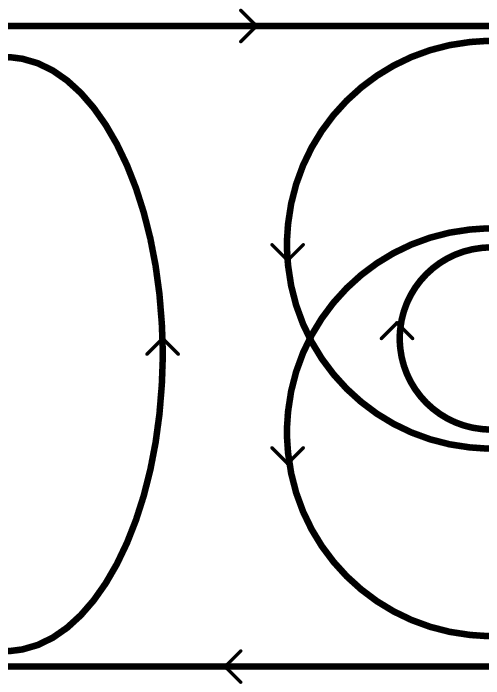}
  \hspace{-6.cm} 
  \begin{minipage}[b]{4cm}
    $1$ \hspace{2cm} $a$

    \vspace{0.8cm}
    $2$

    \vspace{0.7cm}
    $3$

    \vspace{0.7cm}
    $4$ \hspace{2cm} $b$

    \vspace{0.3cm}
  \end{minipage}
  \hspace{1.6cm}
  \begin{minipage}[b]{0.5cm}
    $1$

    \vspace{0.7cm}
    $2$

    \vspace{0.7cm}
    $3$

    \vspace{0.8cm}
    $4$

    \vspace{0.3cm}
  \end{minipage}
  \caption{Examples of a colour flow (left) which contributes in the
    limit of wide angle, hard radiation, and (right) a configuration which
    is suppressed in the same limit. In these diagrams, the final state
    gluons (on the right of each picture) are ordered according to their
    rapidity.}
  \label{fig:Colours}
\end{figure}

The colour-ordered Parke-Taylor amplitudes~\cite{Parke:1986gb} for tree-level
$gg\to g\cdots g$-scattering allow for a very neat
analysis~\cite{DelDuca:1993pp,DelDuca:1995zy} of the dominant colour
configurations in the limit of widely separated, hard gluons. The
conclusion, as presented in Ref.~\cite{DelDuca:1993pp,DelDuca:1995zy}, is
that the leading contribution in this limit of \emph{Multi-Regge-Kinematics}
(\emph{MRK}) is provided by the colour configurations which can be untwisted
to two non-crossing ladders, connecting the rapidity-ordered
gluons. Fig.~\ref{fig:Colours} (left) contains an example of a configuration
contributing in the \emph{MRK}-limit, and one (right) which does not. The
numbering of the final state partons is according to their rapidity.


The colour connections in Fig.~\ref{fig:Colours} (left) can be summarised as
$a134b2a$, and if the point for particle 2 is moved to the left side of the
same plot, then no colour lines cross. This is always possible if in the colour
connection string (like $a134b2a$), the particles entering between the two
initial state gluons are ordered in rapidity, as in the case of
$a134b2a$. 

The colour connections in Fig.~\ref{fig:Colours} (right) can be
summarised as $a1324ba$, which contain an un-ordered string between $a$ and
$b$, and as long as rapidity ordering of the particles is reflected in
the vertical position of the particles in Fig.~\ref{fig:Colours} then one
cannot avoid crossing colour lines by moving the points from the left to the
right side of the plot. This configuration is suppressed in the
\emph{MRK}-limit.

Furthermore, the study of Ref.~\cite{DelDuca:1993pp,DelDuca:1995zy} shows
that all the leading configurations each have the same limit in the
\emph{MRK}-limit, and indeed lead to the colour traces resulting in a colour
factor $\Ca$ for every final state gluon. The limit agrees with that
predicted by the amplitudes of Fadin-Kuraev-Lipatov (FKL)~\cite{Kuraev:1976ge}.

When we pass an event from \emph{High Energy Jets} to \ariadne, we choose a
colour configuration at random from the set of colour connections which are
leading in the \emph{MRK}-limit, and pass the event using an interface conforming
to the \emph{Les Houches accord}~\cite{Boos:2001cv}.

\section{The \protect\ariadne Dipole Cascade}
\label{sec:ariadne}

The \ariadne program \cite{Lonnblad:1992tz} is based on the colour
dipole model developed by the Lund group
\cite{Gustafson:1986db,Gustafson:1987rq,Andersson:1988gp,Lonnblad:1994wk},
where gluon emissions are modelled as coherent radiation from two
colour-connected partons. The general idea is best described in
$e^+e^-$-annihilation into jets, where the inclusive probability of
emitting a gluon from the original $q\bar{q}$-pair is given by the
well known matrix element
\begin{equation}
  \label{eq:ee2qqg}
  D(x_1,x_3)dx_1dx_3=
  \frac{\alphas C_F}{2\pi}\frac{x_1^2+x_3^2}{(1-x_1)(1-x_3)}dx_1dx_3,
\end{equation}
where $x_1$ and $x_3$ are the final-state energy fractions of the
quark and anti-quark respectively after the emission.  A subsequent
gluon emission will then come either from the dipole between the
quark and the gluon or from the one between the gluon and the
anti-quark, with a trivial generalization to further emissions. The
splitting function in eq.~\eqref{eq:ee2qqg} is modified slightly in
the case of dipoles between gluons, giving \eg\ for a gluon--gluon
dipole,
\begin{equation}
  \label{eq:gg2ggg}
  D(x_1,x_3)dx_1dx_3=
  \frac{\alphas N_c}{4\pi}\frac{x_1^3+x_3^3}{(1-x_1)(1-x_3)}dx_1dx_3.
\end{equation}
We note that in the soft and collinear limits we have \eg\ $x_1\to z$,
$x_3\to 1$ and $dx_3/(1-x_3)\to dQ^2/Q^2$, where $z$ and $Q^2$ are the standard energy fraction and virtuality splitting variables, which gives
\begin{equation}
  \label{eq:gg2gggsoft}
  D(x_1,x_3)dx_1dx_3\to
  \frac{\alphas N_c}{4\pi}\frac{1+z^3}{1-z}dz\frac{dQ^2}{Q^2}.
\end{equation}
Since a given gluon splitting gets contributions from two dipoles we
obtain
\begin{equation}
  \label{eq:g2ggsoft}
  D(z,x_3\to1)dzdx_3 + D(1-z,x_3\to1)dzdx_3 =
  \frac{\alphas N_c}{2\pi}\frac{\left(1-z(1-z)\right)^2}{z(1-z)}
  dz\frac{dQ^2}{Q^2},
\end{equation}
and we recover the standard gluon splitting function.

Subsequent emissions are ordered in a Lorentz-invariant transverse
momentum defined as
\begin{equation}
  \label{eq:invpt}
  p_\perp^2=S_{\mathrm{dip}}(1-x_1)(1-x_3)=\frac{s_{12}s_{23}}{S_{\mathrm{dip}}},
\end{equation}
(where $s_{ij}$ is the squared invariant mass of partons $i$ and $j$)
which, together with a conveniently defined rapidity,
\begin{equation}
  \label{eq:invrap}
  y=\frac{1}{2}\ln\frac{1-x_1}{1-x_3}=\frac{1}{2}\ln\frac{s_{23}}{s_{12}},
\end{equation}
results in a dipole splitting function which can be approximated by
\begin{equation}
  \label{eq:dipsplit}
  D(p_\perp^2,y)dp_\perp^2dy\propto\frac{dp_\perp^2}{p_\perp^2}dy.
\end{equation}
The emissions are then made exclusive by introducing no-emission
probabilities, or Sudakov form factors, giving the probability that
there were no emissions beteen two scales,
\begin{equation}
  \label{eq:sud}
  \Delta(p_{\perp1}^2,p_{\perp2}^2)=\exp\left(-\int_{p_{\perp1}^2}^{p_{\perp2}^2}
    dp_\perp^2\int dy D(p_\perp^2,y)\right)
\end{equation}
giving rise to the standard (next-to-leading) logarithmic resummation
of soft and collinear divergences.

In the final state radiation, \ariadne also includes the $g\to
q\bar{q}$ splitting, but in this paper we will only concern ourselves
with gluon emissions and we will therefore not go into further details.

\ariadne has a fairly unique way of handling radiation in collisions
where there are incoming hadrons. In a normal parton shower one would
apply a backwards evolution of initial-state splittings, and in more
recent dipole shower implementations such as those in \pythia
\cite{Sjostrand:2007gs} and \sherpa \cite{Gleisberg:2008ta}, dipoles
are defined between incoming and outgoing partons in the hard
interaction. The \ariadne program, in contrast, uses the so-called
Soft Radiation Model \cite{Andersson:1988gp}, where there are dipoles
between the hadron remnants and the partons from the hard
interactions.

Here we will rely on the \HEJ program to generate the initial-state
emissions, and we will therefore not go into details of this Soft
Radiation Model. Instead we will go back to the dipole splitting
function in \refeq{eq:dipsplit} to see how it relates to the matrix
elements generated in \HEJ.

The standard $g\to gg$ splitting function can be derived from ratios
of matrix elements (see \eg\ chapter 5 in \cite{Ellis:1991qj}) as
\begin{equation}
  \label{eq:splitggg}
  d\sigma_{n+1}=
  \frac{\left|{\cal M}_{n+1}\right|^2}{\left|{\cal M}_{n}\right|^2}
  \frac{dk_\perp^2dz}{16\pi^2}d\sigma_n\approx
  \frac{\alphas}{2\pi} P_{gg}(z)\frac{dk_\perp^2}{k_\perp^2}dzd\sigma_n.
\end{equation}
Looking at the dipole splitting function in \refeq{eq:dipsplit}, we can
associate the emitted gluon to one or the other emitter, depending on which is
closer. We can define \eg\ $1-z=x_1/(2-x_3)$ and in the limit where parton 3
retains most of its energy we have $dy\approx dz/z$ giving in the end
\begin{equation}
  \label{eq:dipsplitme}
  D(p_\perp^2,y)\approx\frac{z}{16\pi^2}
  \frac{\left|{\cal M}_{n+1}\right|^2}{\left|{\cal M}_{n}\right|^2}.
\end{equation}
This relates the matrix elements to splitting functions, and we will utilise
this to define a subtraction term for the shower, based on the matrix
elements used in \HEJ. With this we can proceed with subtracting the
radiation probabilities arising in \HEJ from the \ariadne dipole splittings.

\section{The Subtracted Shower}
\label{sec:matching}
The idea for combining the all-order resummations of \HEJ and \ariadne is to
first let \HEJ generate an event according to
Eq.~(\ref{eq:resumdijetFKLmatched}), and then let \ariadne shower these
events using a splitting function, which has been modified to subtract the
effects of the resummation already included in \HEJ.

The events from HEJ will consist of parton configurations where the two
partons extremal in rapidity are required to be members of hard jets (where
the scale of hardness is chosen in the analysis) with absolute rapidities less than a cutoff $y_\mathrm{max}$ of, say, 5.5. The
minimum transverse momenta of these extremal partons can be required to be
larger than a scale not much smaller than the jet scale used in the
analysis~\cite{Andersen:2011hs}. The phase space in-between these two extremal
partons is populated by gluons with transverse momenta above some small
cut-off $\lambda$ (of order 1~GeV), below which a subtraction is applied, in
order to organise the cancellation of IR divergences between real and virtual
corrections. The \HEJ events therefore consist of partons of any transverse
momentum, and with absolute rapidities less than $y_\mathrm{max}$, with the
transverse momentum of the two extremal partons larger than the minimum jet
transverse scale.

These events are then given to \ariadne. When \ariadne performs a trial
emission, we subtract from the \ariadne splitting function the effective
splitting function as obtained from \emph{HEJ} for the one extra
\ariadne-emission, as indicated in Eq.~(\ref{eq:dipsplitme})
\begin{align}
  D_\mathrm{subt}(p_\perp^2,y)=\frac z{16\pi^2}\ \frac{|\mathcal{M}_{n+1}^\mathrm{t}|^2}{|\mathcal{M}_n^\mathrm{t}|^2},
  \label{eq:HEJsubtract}
\end{align}
where the matrix elements are evaluated using the tree-level
\emph{HEJ}-formalism of Eq.~(\ref{eq:multijetVs}) (with matching to
high-multiplicity full tree-level matrix element, if appropriate). The ratio
of the square of matrix elements in Eq.~(\ref{eq:HEJsubtract}) is given by
the factors included in the brackets on the last line of
Eq.~(\ref{eq:multijetVs}) --- up to the effects of the momentum reshuffling
in order to keep the incoming momenta on-shell after the trial emission
of the shower (see below). However, in order to take properly into account
the effects of the reshuffling, we evaluate the full matrix element (not just the
extra factors of $V^\mu V_\mu$) on the respective momenta. The square of the
matrix elements is Lorentz-invariant, and the value of $z$ is retained from
\ariadne in the proper dipole frame. This method therefore avoids the need
for any Jacobians from a change of reference system.

The $n$-particle configuration is the original \emph{HEJ}-event, and
the \ariadne trial emission is added to the final state configuration for the
evaluation of the matrix element for the $n+1$-particle configuration with a
\emph{recoil strategy}, which keeps the incoming partons with zero transverse
momentum:
\begin{enumerate}
\item The transverse momentum of the trial emission is subtracted from the
  other final state partons, with the subtraction distributed proportional to
  the transverse momentum of each parton.
\item The new energy and longitudinal momentum of each parton is defined to
  keep fixed the rapidity of each of the original partons.
\end{enumerate}
Various algorithms were devised for dividing the subtraction of the momentum
of the trial emission onto the original \HEJ partons, each giving very
similar results for all the distributions and observables tested. We also
tested the subtraction using both the tree-level and the fully regulated
formalism for \HEJ, and the differences are minimal (as expected, since the
difference is higher order effects). We therefore choose to use just the
tree-level matrix elements, since they evaluate slightly faster (and one
avoids complications from the reshuffling moving momenta across the
regulating parameter $\lambda$).

No subtraction is applied if the \ariadne trial emission is outside the phase
space where \HEJ would emit gluons, \eg~if the trial emission would result
in a configuration where the most forward or backward gluon is softer than
the jet scale. This region of emissions collinear to the incoming beam is a
domain which is populated purely by the shower.

The described procedure ensures that all the emissions performed by
\ariadne have been properly subtracted to avoid double-counting. But
it should also be noted that the veto algorithm\footnote{See \eg\ the
  description of the veto algorithm in the appendix of
  \cite{Buckley:2011ms}.} used to order the emissions in \ariadne,
automatically ensures that only the subtracted splitting kernel is
exponentiated in the Sudakov form factors.

The ordering of emissions in (invariant) transverse momentum in
\ariadne is of some concern. As \HEJ produces also some very soft
gluons ($\sim1$~GeV), there may arise situations where a soft gluon
ends up fairly close to a hard one. In this situation \ariadne would
normally first emit some hard, collinearly enhanced, emissions before
the softer ones, but now, the dipole between the soft and hard gluon
from \HEJ becomes too small to allow the full range of
collinearly-enhanced radiation. The effect is that the very hard jets
in our matching procedure will become a bit too narrow. This issue
will be discussed further when we look at the resulting jet shapes in
the next section.

\subsection{The Algorithm}
\label{sec:algorithm}

What follows here is a step-by-step description of the algorithm as
implemented in the \HEJ/\ariadne interface.

\begin{enumerate}
\item Generate a partonic state with \HEJ, using a given cutoff in
  transverse momentum for the extremal partons, and a small transverse
  momentum cutoff for soft gluons.
\item This state is then sent via \pythia to \ariadne using the Les
  Houches interface.
\item \ariadne then sets up its internal dipole event record and
  starts the dipole cascade starting from a scale given by the largest
  transverse momentum of any of the \HEJ partons.
\item For each potential emission in the dipole cascade, check whether
  it corresponds to something that could have been produced by \HEJ,
  i.e.\ a gluon emission with a rapidity between the two extremal
  jets. If this is the case:
  \begin{itemize}
  \item Calculate the kinematics of the gluon and the approximate
    ratio of matrix elements corresponding to the splitting function
    used $r_A=\left|{\cal M}_{n+1}\right|^2/\left|{\cal
        M}_{n}\right|^2$ according to Eq.~(\ref{eq:dipsplitme}).
  \item This gluon is then sent to \HEJ using a call-back function
    where it is inserted in the original partonic state and calculate
    the ratio of the corresponding matrix element and the original
    matrix element, $r_H=\left|{\cal M}^t_{n+1}\right|^2/\left|{\cal
        M}^t_{n}\right|^2$.
  \item Veto the emission with probability $r_h/r_a$. This is
    equivalent to using the subtracted splitting function
    $D(p_\perp^2,y)- D_\mathrm{subt}(p_\perp^2,y)$ in the shower. In
    addition as this is done inside the veto algorithm, it is the
    subtracted splitting function which is resummed in the Sudakov
    form factor.
  \end{itemize}
  If the emission is outside the extremal jets, it is kept only if it
  has a transverse momentum below the cutoff for extremal jets.
  Furthermore if the emission is a final-state splitting of a gluon
  into a $q\bar{q}$ pair, it is simply kept as such emissions cannot
  be produced by \HEJ. Note also that any gluon emission with a transverse
  momentum below the
  phase space slicing parameter $\lambda$ in \HEJ will be kept.
\item After the dipole cascade in \ariadne, the partonic final state
  is hadronized by \pythia, where also the decay of unstable hadrons
  is performed.
\end{enumerate}

A few notes are in order.
\begin{itemize}
\item First it is clear that we avoid double-counting all over phase
  space. For each emission in the shower, only those which could not have
  been produced in \HEJ are kept without change. For those
  which do correspond to a \HEJ emission, the corresponding emission
  probability is subtracted, basically only retaining the collinear
  pole, while the soft gluons (above the \HEJ cutoff) are effectively vetoed
  by the same subtraction algorithm.
\item The emissions which are affected by the Soft Radiation Model in
  \ariadne are those related to dipoles connecting the proton
  remnants, and these will give gluons outside the extremal jets, and
  are therefore forced to be below the extremal jet cut. Hence,
  possible effects of the unconventional treatment of initial state
  radiation in \ariadne are minimized, and will not lead to extra jets.
\item It can be argued that if a gluon emission which is accepted in
  \ariadne corresponds to a partonic configuration which is not
  possible to produce by \HEJ, further emissions from the
  dipoles connected to this gluon should be unrestricted. Clearly if
  the gluon was collinear to a \HEJ-jet, the emissions from the dipole
  between that jet and the gluon should not be subtracted. On the
  other hand, emissions from the other dipole should still be
  subtracted, and the subtraction in the first dipole should be small
  since it would always correspond to collinear emission. We have
  tried two scenaria: one where all emissions are subtracted, and one where only emissions from dipoles between
  original \HEJ partons are subtracted (this is
  the default). The differences turned out to
  be almost negligible.
\end{itemize}

\section{Results}
\label{sec:results}

In this section we compare first the effective splitting function from \HEJ
with that of \ariadne, in both the soft and the collinear regions. We then
study in details the effects of the shower on a sample event-configuration
from \HEJ, before moving on to a study of the resulting shower profiles,
which are compared to \atlas data. Finally, we discuss the impact of
the shower on a few observables discussed in
\eg~Ref.~\cite{Andersen:2003gs,Andersen:2008gc,Andersen:2010ih,Andersen:2011hs},
which are sensitive to the description of hard, radiative corrections.

\subsection{Comparison of Splitting Functions}
\label{sec:comp-splitt-funct}

\begin{figure}[tbp]
  \centering
  \epsfig{width=.8\textwidth,file=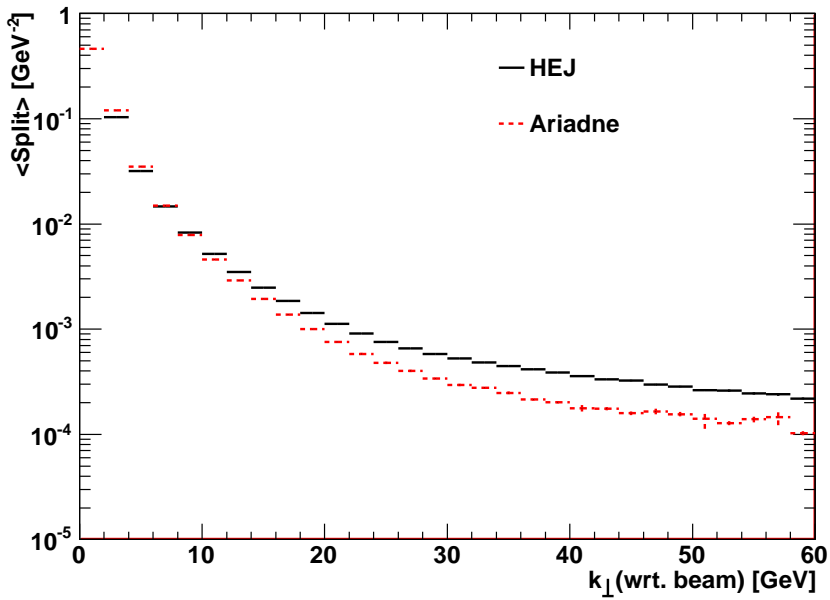}
  \caption{The average value of $D/z$ as a function of the transverse
    momentum of the trail splitting, for the subset of wide-angle emissions
    discussed in the text. Hard, wide-angle emissions from \ariadne are (on
    average) vetoed, since the effective splitting function from \HEJ is (on
    average) larger than that from \ariadne.}
  \label{fig:softsplit}
\end{figure}
In Fig.~\ref{fig:softsplit} we compare the average value of the splitting
functions $D/z$ calculated in \ariadne and with \HEJ
(Eq.~(\ref{eq:HEJsubtract})), after division with \as (to remove the effects
of a different evaluation of \as in \ariadne and \HEJ), as a function of the
transverse momentum of the trial emission. The average is over each bin in
the distribution, for an unweighted HEJ event sample of 26~GeV dijets, where
at least one reconstructed jet above 30~GeV is required in the post-shower
analysis. We have used only the subset of \ariadne trial splittings where
the original dipole consists of partons from \HEJ, and where the trial
splitting is at least a distance $R=0.5$ away from any of the original \HEJ
partons. This is to test only the soft (but not collinear) description in the
two frameworks. Generally, the effective splitting functions for wide-angle
emissions are quite similar in \HEJ and \ariadne. However, we see that, on
average, the \ariadne splitting function is larger than the effective
splitting function from \HEJ only for emissions of transverse momenta less
than about 10~GeV, and then only by a small amount. This means that
effectively, wide-angle emissions harder than about 10~GeV are automatically
vetoed in the subtraction mechanism, since the probability for emissions is
larger in \HEJ than in \ariadne.

We see by comparing the explicit numbers that the \HEJ splitting function
(after division by \as) tends to $\frac{\Ca}{\pi k_\perp^2}$ in the MRK
region of semi-hard, wide-angle emissions. This is the result of the full
tree-level QCD (and the pure BFKL formalism).

\begin{figure}[tbp]
  \centering
  \epsfig{width=.8\textwidth,file=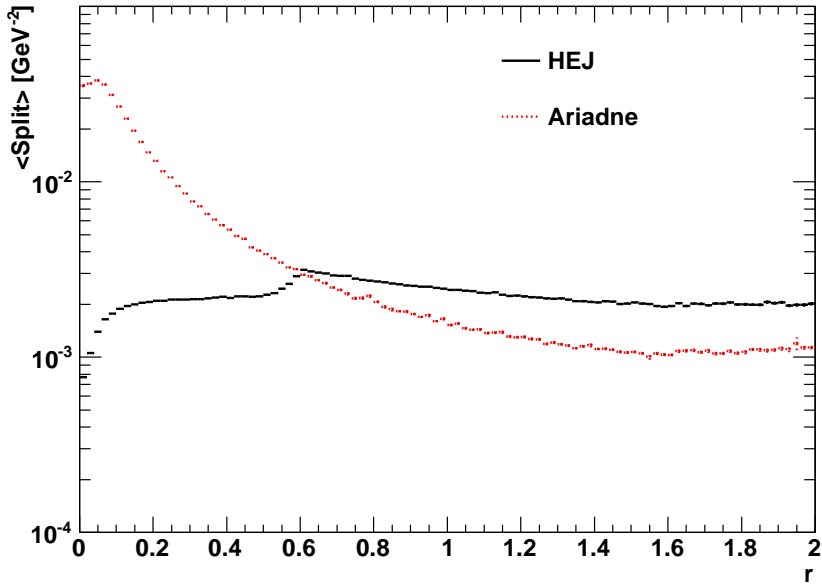}
  \caption{The average value of $D/z$ for trial splittings, as a function of
    the distance $r$ from the \HEJ partons, for emissions with transverse
    momentum greater than 10~GeV. In the collinear region, the \ariadne
    splitting function is much larger than the subtraction from \HEJ, whereas
  \HEJ dominates at larger values of $r$.}
  \label{fig:collsplit}
\end{figure}

In Fig.~\ref{fig:collsplit} we compare the average value of $D/z$ in \ariadne
and that arising from \HEJ for a sample of \ariadne trial emissions (again
from original \HEJ partons) of harder than 10~GeV in transverse momentum, as
a function of the distance $r$ (in (rapidity,$\phi$)) to the nearest \HEJ
parton. Here we see that at small $r$, the \ariadne splitting function is an
order of magnitude larger than the subtraction term from \HEJ. The \HEJ
subtraction term is (on average) larger than the (average) \ariadne splitting
function only for $r>0.6$, which in this case was also chosen as the jet size
parameter in the anti-kt jet clustering. The change in behaviour for the
effective \HEJ splitting function at the jet-size parameter is because of the
use of two different effective emission vertices, depending on whether or not
the additional \HEJ-emissions are collinear to the partons extremal in
rapidity.

\subsection{The Description of Jet Structure}
\label{sec:descr-jet-struct}

Based on the analysis of the previous subsection, one would expect that the
effect of the \ariadne showering of the \HEJ events would be to radiate
mostly in the immediate surroundings of the existing \HEJ partons, whereas hard
wide-angle emissions from \ariadne should be suppressed.

\begin{figure}
  \centering
  \epsfig{width=.8\textwidth,file=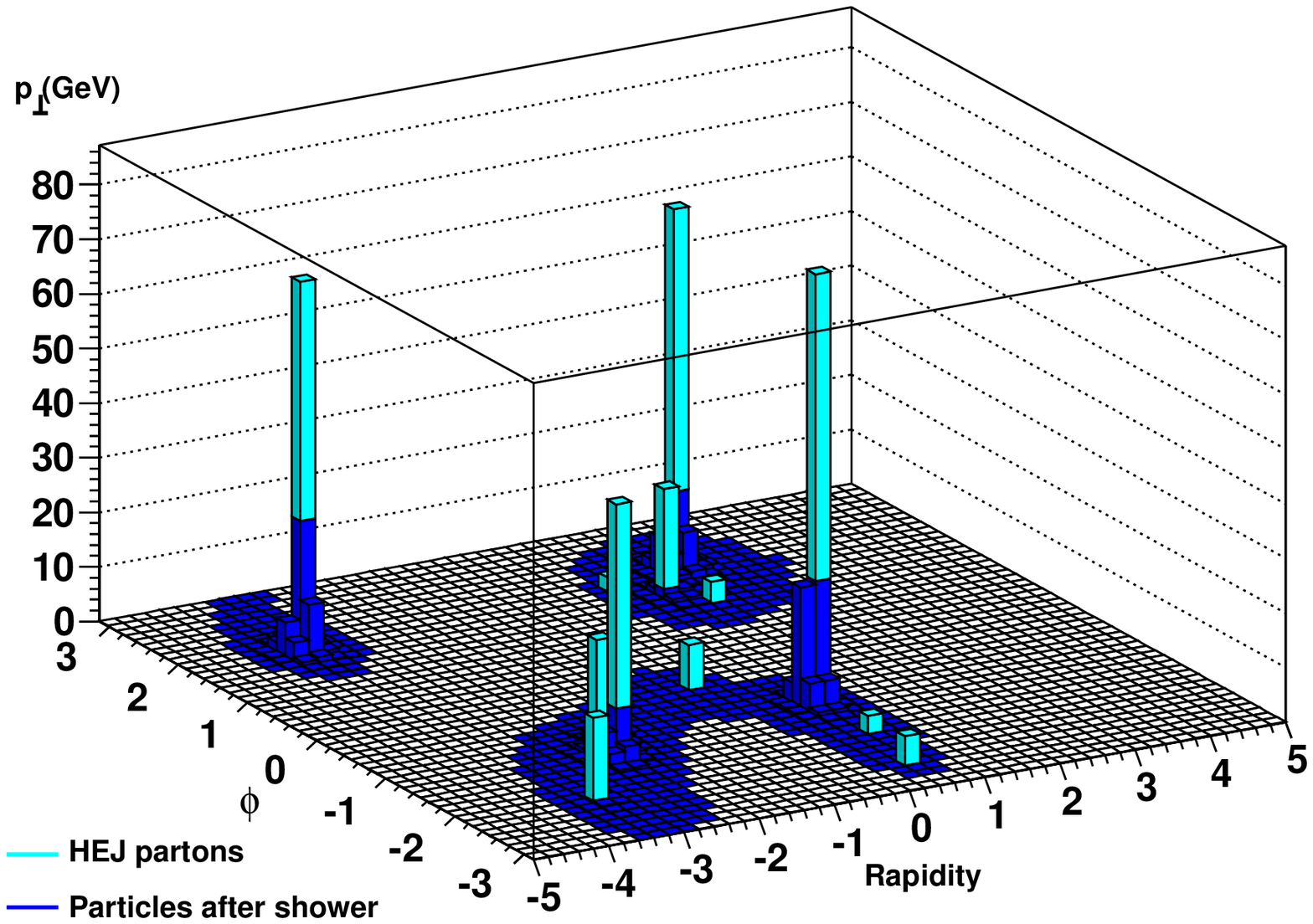}

  \epsfig{width=.49\textwidth,file=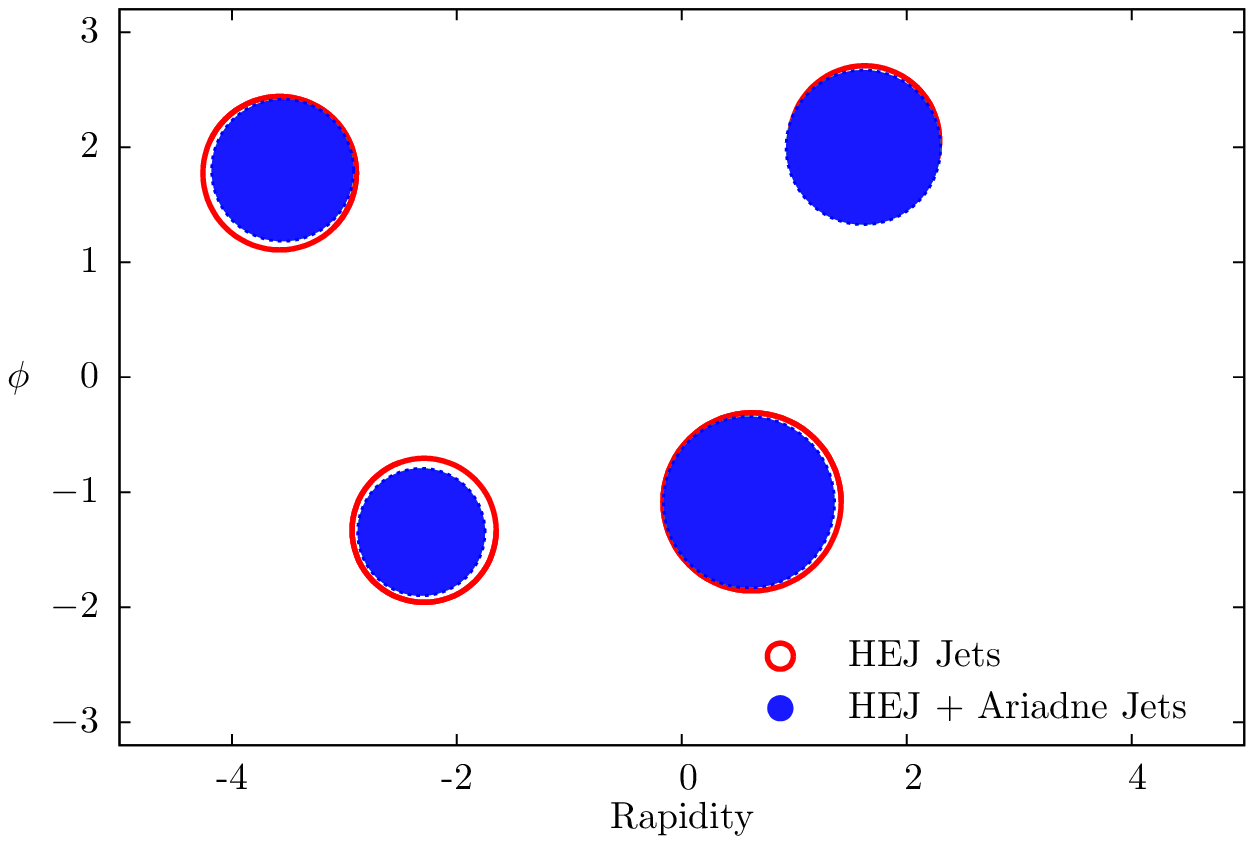}
  \epsfig{width=.49\textwidth,file=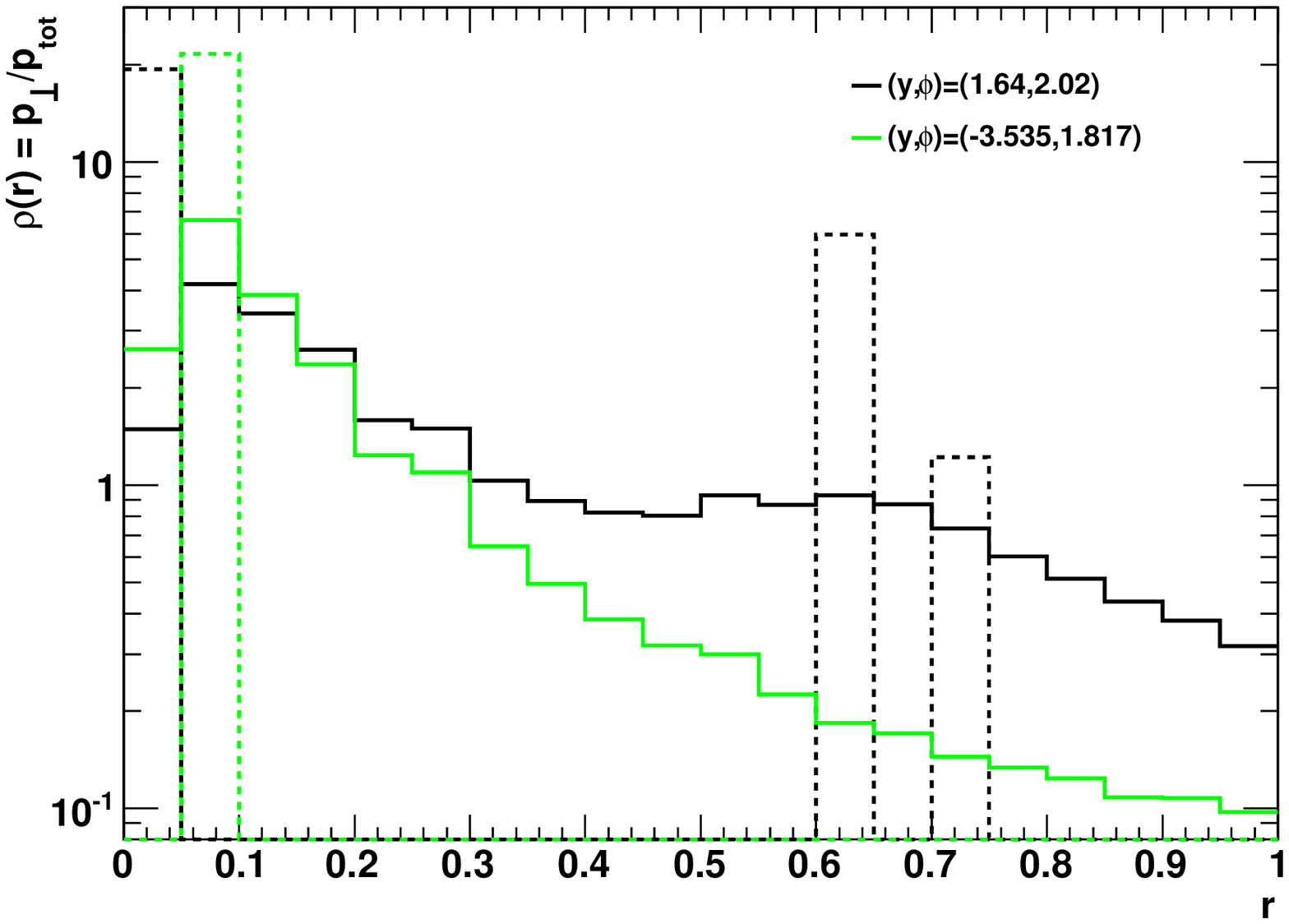}
  \caption{Top: A Lego$^{\textregistered}$-plot of the momentum configuration
    from HEJ, and the average outcome from 10000 showers of the same
    HEJ-event.  Bottom Left: the 4 hard ($p_\perp>60$~GeV) jets resulting from the
    jet clustering, before and after the showering.  The radii of the circles
    are proportional to the momentum of the jet.  Bottom right: The shower
    profile of two of the jets, as a function of $r=\sqrt{\Delta
      y^2+\Delta\phi^2}$.  The dotted lines correspond to the initial partons in
  each case.}
  \label{fig:lego}
\end{figure}
Fig.~\ref{fig:lego} illustrates the effects of the shower on a sample event from
\HEJ, where the average outcome of 10,000 showers is plotted along with the
initial partons.  The top Lego plot illustrates clearly the smearing effect of
the shower.  The effect on the momentum of the reconstructed hard jets (defined
with the anti-$k_\perp$ algorithm and $R=0.6$ and with a transverse momentum
larger than 30~GeV) is shown in the bottom left plot, where the radii of the
circles is proportional to the momentum of the reconstructed jets.  After
showering, the jets move only a modest amount in $(y,\phi)$ and there is only a
small change in the momentum.  In this particular sample event, the momenta of
three of the jets decreases slightly (from \{64, 68, 79\}~GeV to \{57, 63,
76\}~GeV respectively) while the momentum of the other one increases very
slightly from 67~GeV to 69~GeV.  Further details can be seen in the jet profile
plot in the bottom right of Fig.~\ref{fig:lego}, where the variable plotted is
the fraction of the jet's transverse momentum found at a radius $r$ from the jet
center:
\begin{equation}
  \label{eq:rhoofr}
  \rho(r)=\frac{1}{p_{\perp}(R)}\frac{dp_\perp(r)}{dr},
\end{equation}
where $p_\perp(r)$ is the summed transverse momentum in bins of radius $r$ and
$R$ is the jet radius used in the anti-$k_\perp$ algorithm
\cite{Cacciari:2008gp}.

\begin{figure}[htbp]
  \centering
  \epsfig{width=.95\textwidth,file=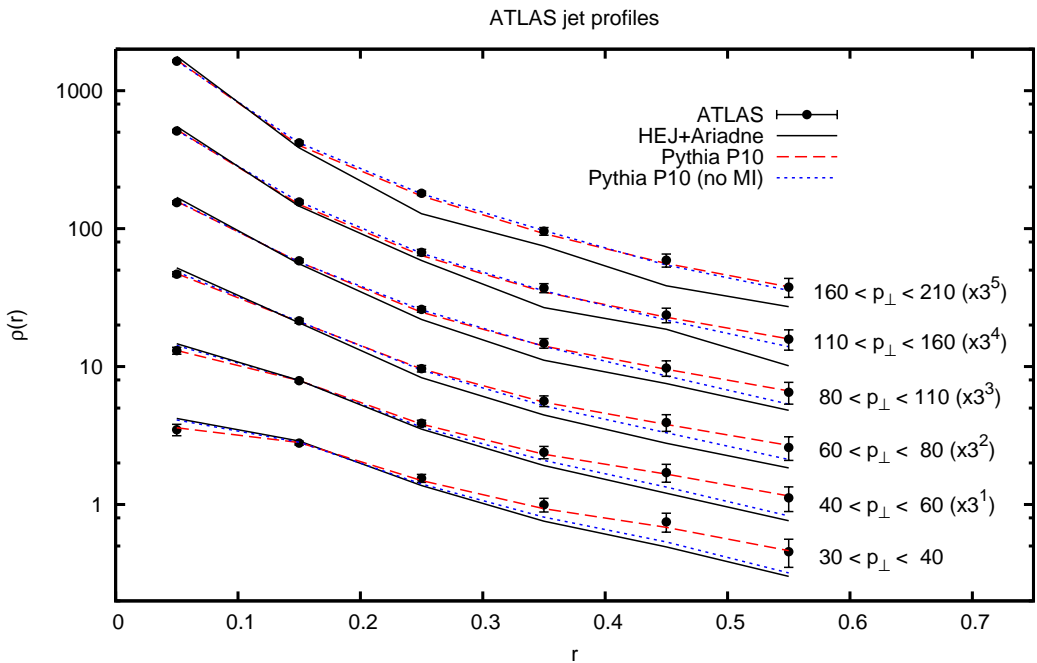}
  \caption{\label{fig:atlasprofile}Jet profiles as measured by \atlas
    \cite{Aad:2011kq} for different bins of jet transvese momenta (in
    GeV) compared with our results (for $\lambda=2$~GeV) (full
    lines). Also shown are the results from \pythia Perugia 10 tune
    with (dashed) and without (dotted) multiple interactions.}
\end{figure}

In Fig.~\ref{fig:atlasprofile} we present our fully simulated pure QCD
inclusive one-jet
events (requiring at least two jets of 26~GeV from HEJ, at least one 30~GeV jet
after showering) compared to a recent measurement of jet shapes by the \atlas
collaboration \cite{Aad:2011kq}. Note that our program does not include any
underlying event simulations. We therefore also compare our results with a
recent tuning of the \pythia program \cite{Sjostrand:2006za} (Perugia 10
\cite{Skands:2010ak}) with and without multiple interactions to estimate such
effects. As expected, the underlying event mainly contributes to jets with small
transverse momenta in the outskirts of the jets.

We find that our jets are very similar to the \pythia ones (without multiple
interactions) for small transverse momenta, which is a good indication that our
matching works as expected. However, at larger $p_\perp$ our jets seem to be a
bit too narrow as compared to data and \pythia. The reason for this is the very
soft gluons fairly close to the hard ones, which are included in the \HEJ
resummation. Such situations result in dipole masses which are too small to
allow for enough radiation from the hard gluons within \ariadne.

The problem is that the cascade in \ariadne becomes somewhat un-ordered --- the
soft gluons should normally be radiated \emph{after} the harder collinear ones
as noted in the previous section. This is a problem also in other matching
procedures, as noted in \cite{Lavesson:2007uu}, especially for the so-called MLM
\cite{MLM,Mangano:2001xp} and PseudoJet \cite{Mrenna:2003if} algorithms. Indeed,
in \cite{Aad:2011kq} the \atlas collaboration see a tendency in the \alpgen
generator, which uses MLM-merging, to produce too narrow jets at large
transverse momenta.

We have here run with a lower cutoff (or regularisation parameter) of the transverse momentum in \HEJ of $2$~GeV, and we have noted
that the effect is enhanced if this cutoff is lowered further. Increasing the
regularisation parameter further will lead to unacceptably large
cancellations between positive and negative weight events. Negative events do
not arise in \HEJ when the regularisation parameter is allowed to be much
smaller. To increase the
cutoff above $2$~GeV is therefore numerically unacceptable, and in any case
it would only reduce the effect, not remove it completely.

One could debate whether the subtraction should be applied to \emph{all}
\ariadne trial emissions, or only to emissions from \ariadne dipoles made up
of partons from \HEJ. In practice this makes very little difference to the
results -- certainly, the difference between the two choices are inseparable
when presented as in Fig.~\ref{fig:atlasprofile} (and
Figs.~\ref{fig:AvgJets}--\ref{fig:gapfraction}). We therefore choose to apply
the subtraction only in the first emission, in order to limit the amount of
momentum reshuffling performed.

\subsection{Impact on Multi-Jet Observables}
\label{sec:impact-multi-jet}

\begin{figure}
  \centering
  \epsfig{width=.8\textwidth,file=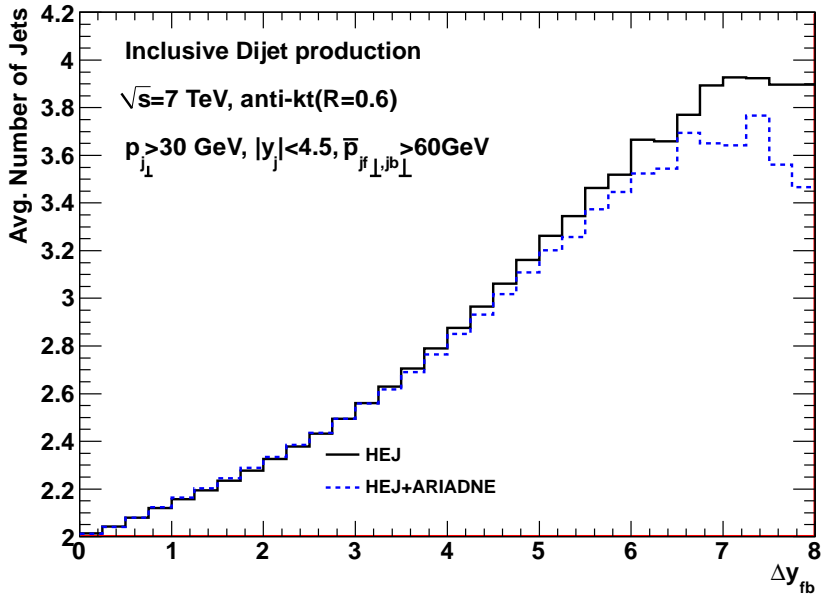}
  \caption{The average number of jets vs.~the rapidity difference between the
  most forward/backward jet as obtained with HEJ, and with HEJ+\ariadne. The
  total effects of shower and hadronisation on this observable are minor.}
  \label{fig:AvgJets}
\end{figure}
Finally, we are ready to study the impact on a few observables of the
addition of a shower to the resummation of \HEJ. We choose to do so for the
observables from a study by \atlas~\cite{Atlas:2010xx}, for which the
prediction from \HEJ were presented in Ref.~\cite{Andersen:2011hs}. We do not
here present a full analysis of the uncertainties from scale and
pdf-variation, as performed in Ref.~\cite{Andersen:2011hs}, but choose to use
as factorisation and renormalisation scale the maximum transverse momentum of
any jet, and include the $\beta_0\log$-terms, as discussed in
Ref.~\cite{Andersen:2011hs,Andersen:2003an,Andersen:2003wy}.

In Fig.~\ref{fig:AvgJets} we compare the average number of jets in a
dijet-sample (anti-kt,
$R=0.6$) with a transverse momentum greater than 30~GeV (and rapidity less
than 4.5) as a function of the rapidity difference between the most forward
and backward hard jet. These two jets are furthermore required to have an
average transverse momentum greater than 60~GeV. The black line is the
partonic prediction from \HEJ (based on just the FKL-configurations discussed
in the current paper). The blue dashed line is obtained after the further
shower and hadronisation by \ariadne. The changes (due to the showering) in
the number of hard jets is very small indeed, but increasing in significance
with the rapidity span and the average number of jets. At a rapidity span of
6 units, the showering and hadronisation leads to a reduction in the average
number of hard jets from roughly 3.6 to slightly above 3.5. For small
rapidity spans ($\Delta y<2$), the showering and hadronisation leads to a
minute increase in the average number of hard jets, whereas for larger
rapidity spans the effect is a larger decrease.

\begin{figure}
  \centering
  \epsfig{width=.8\textwidth,file=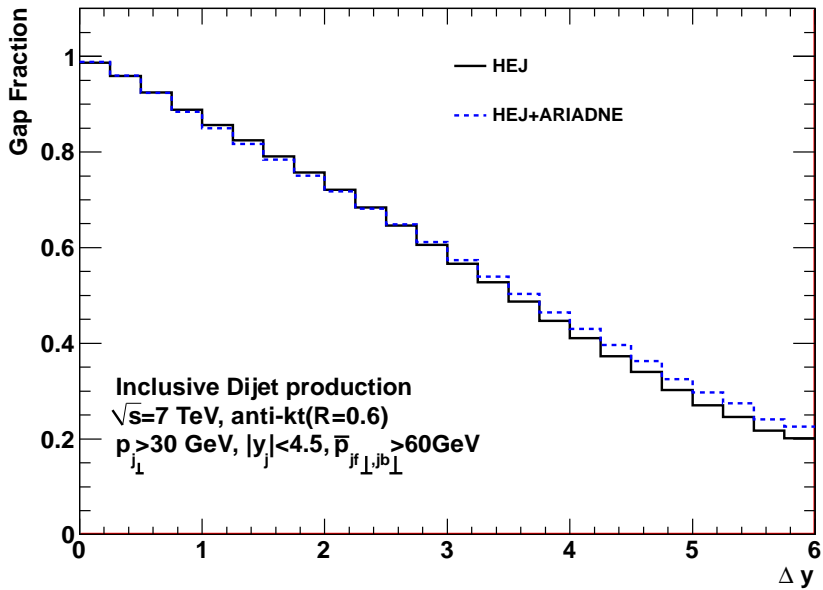}
  \caption{The gap fraction (exclusive dijet over inclusive dijet rate) as a
    function of the rapidity span between the most forward/backward hard jet,
    as obtained with HEJ, and with HEJ+\ariadne. The total effects of shower
    and hadronisation on this observable are minor.}
  \label{fig:gapfraction}
\end{figure}
The trend and small effect is found again in the prediction of the \emph{gap
  fraction} in Fig.~\ref{fig:gapfraction}, defined here as the exclusive
dijet rate over the inclusive dijet rate, as a function of the rapidity span
between the most forward/backward pair of hard jets ($p_\perp>30$~GeV). The
effects seen in Fig.~\ref{fig:AvgJets} are repeated here -- for small rapidity
spans ($\Delta y<2$), the showering and hadronisation leads to a minute
decrease in the gap fraction, but for larger rapidity spans, the effect is
a small increase.

\section{Outlook}
\label{sec:outlook}

We have here presented a procedure for obtaining exclusive hadronic final
states within the framework of \HEJ, describing jet production in high-energy
hadronic collisions. To do this we have developed a new kind of matching
scheme dealing with two separate all-order summations, where the hard jets
(and soft contributions) are first generated in \HEJ, followed by a
final-state shower dressing up the jets with further subtracted soft and
collinear radiation. Double-counting is avoided by carefully removing the
appropriate soft divergences in the parton shower in a way such that the
collinear divergences are still correctly exponentiated.

We have validated our procedure both technically, by making sure that
the subtracted splitting functions look reasonable, and also by
comparing the resulting jet shapes with recent measurements by the
\atlas experiment. For the latter we find very good agreement with
data. We have also shown that previous predictions published for the
\HEJ model at parton level are fairly insensitive to the addition of
parton showers and hadronization for reasonable choices of jet
definitions.

We have, however, found that our results are somewhat sensitive to the very
soft gluon emissions also included in \HEJ. The reason can be understood from
the fact that \ariadne is forced to work in an un-ordered way. Some soft
emissions from \HEJ, would normally be emitted by \ariadne after the emission
in the collinear region. This indirectly reduces the phase space available
for collinear emissions, giving jets which are a bit too narrow, especially
at high transverse momenta. The situation is similar to the problems found in
\cite{Lavesson:2007uu} when investigating the so-called MLM
\cite{MLM,Mangano:2001xp} and PseudoJet \cite{Mrenna:2003if} merging
procedures.

In the future we will investigate different ways of solving this
problem, and we have already identified two different solutions.  One
is based on the CKKW(-L) \cite{Catani:2001cc,Lonnblad:2001iq} merging
procedure. Here the states produced by \HEJ would first be reweighted
by Sudakov form factors produced by the properly subtracted parton
shower, whereafter the shower can be added in the same way as
presented in this article. Here we plan to use a recent CKKW-L
implementation in the \pythia generator \cite{Lonnblad:2011xxxx}, which will
also allow us to add underlying events from the multiple-interaction
model in \pythia.

Another option we will pursue is related to the concept of
\textit{primary} or \textit{back-bone} gluons introduced in
\cite{Andersson:1995ju} and \cite{Salam:1999ft}. There it is found
that the main real-emission contribution to the cross section
according to BFKL is given by gluons which are ordered in both
positive and negative light-cone momenta, while other emissions can be
conveniently summed over. In our case that would mean that soft gluons
produced in \HEJ that are not ordered in this fashion would simply be
removed in a kind of \textit{post-factum} resummation. After that the parton
shower, this time \emph{unsubtracted}, can be applied with a simple phase
space veto, allowing only gluons emissions which are un-ordered in
light-cone momenta, in the same way as has been done in
\cite{Kharraziha:1997dn} and \cite{Flensburg:2011kk}.

Nevertheless, we have found that the effects of the non-ordering of
the parton shower are small, and the current implementation is
therefore an important step forward in the description of multi-jet events in
hadronic collisions. This is the first framework that will properly
simulate complete multi-jet final states with proper resummation of also
semi-hard emissions at high energies, and will provide an important
alternative to the conventional event generator approaches, which are all
based on DGLAP resummation only.

\section*{Acknowledgments}

We would like to thank Frank Krauss, Gavin Salam, Peter Richardson, Mike
Seymour and Peter Z.~Skands for discussions on matching to parton showers.

Work supported in part by the EU Marie Curie RTN MCnet
(MRTN-CT-2006-035606), the Swedish research council (contracts
621-2008-4252 and 621-2009-4076) and the UK Science and Technology Facilities
Council (STFC).

L.L.~gratefully acknowledges the hospitality of the CERN theory unit. JRA
also acknowledges the support of CERN TH during various stages of this work.

\bibliographystyle{JHEP}  
\bibliography{refs} 

\end{document}